\documentclass[twocolumn]{aastex63}
\usepackage{newtxtext,newtxmath}
\usepackage[T1]{fontenc}
\usepackage{ae,aecompl}
\usepackage{graphicx}	
\usepackage{amsmath}	
\usepackage{amssymb}
\usepackage{float}
\usepackage[flushleft]{threeparttable}%to add a comment under tables
\usepackage{changepage}
\usepackage{epsf}
\usepackage{placeins}

\shorttitle{Catastropphic Outlier Photo-$z$ Mitigation}
\shortauthors{Wyatt \& Singal}
\reportnum{Pub. Astron. Soc. Pacific}

\begin{document}

\title{Outlier Prediction and Training Set Modification to Reduce\\Catastrophic Outlier Redshift Estimates in Large-Scale Surveys}

\author{M. Wyatt and J. Singal}
\affiliation{Physics Department, University of Richmond\\138 UR Drive, Richmond, VA 23173, USA}

\email{michael.wyatt@physics.ucla.edu}

\begin{abstract} 
We present results of using individual galaxies' probability distribution over redshift as a method of identifying potential catastrophic outliers in empirical photometric redshift estimation. In the course of developing this approach we develop a method of modification of the redshift distribution of training sets to improve both the baseline accuracy of high redshift ($z >$ 1.5) estimation as well as catastrophic outlier mitigation. We demonstrate these using two real test data sets and one simulated test data set spanning a wide redshift range (0 $< z <$ 4). Results presented here inform an example `prescription' that can be applied as a realistic photometric redshift estimation scenario for a hypothetical large-scale survey. We find that with appropriate optimization, we can identify a significant percentage ($>$30\%) of catastrophic outlier galaxies while simultaneously incorrectly flagging only a small percentage ($<$7\% and in many cases $<$3\%) of non-outlier galaxies as catastrophic outliers. We find also that our training set redshift distribution modification results in a significant ($>$10) percentage point decrease of outlier galaxies for $z >$ 1.5 with only a small ($<$3) percentage point increase of outlier galaxies for $z <$ 1.5 compared to the unmodified training set. In addition, we find that this modification can in some cases cause a significant (\textasciitilde 20) percentage point decrease of galaxies which are non-outliers but which have been incorrectly identified as outliers, while in other cases cause only a small ($<$1) increase in this metric.
\\
\end{abstract}

\keywords{galaxies: statistics -- methods: miscellaneous -- techniques: photometric\\}

\section{Introduction} \label{intro}
In the near future, large scale surveys such as the Rubin Observatory Legacy Survey of Space and Time \citep[LSST -- e.g.][]{Ivezic08} will observe up to hundreds of millions of individual galaxies in a limited number of photometric bands. One important quantity that must be determined from the data collected in these surveys is each galaxy's redshift, which is a key measurement for almost every science goal in extragalactic astrophysics and cosmology. 

\begin{figure} 
\begin{center}
\hspace*{-0.14in}
\includegraphics[width=3.00in]{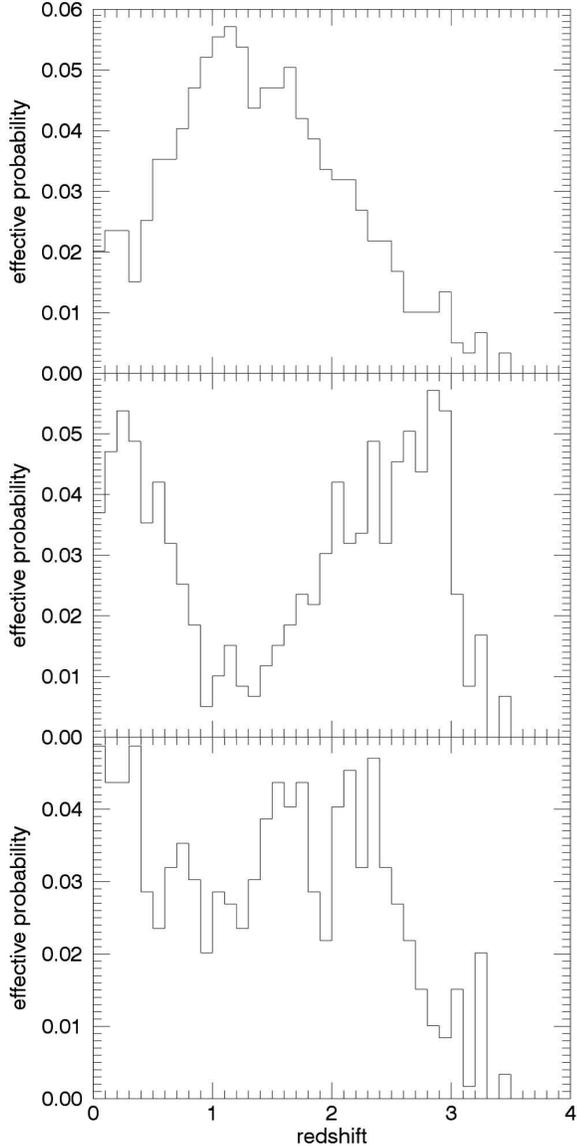}
\caption{Three examples of individual real galaxy EPDFs (effective probability vs. redshift) output by a SPIDERz photo-$z$ determination using the COSMOS-reliable-$z$ test data set described in \S \ref{tdata} with 2000 training galaxies. The top panel EPDF manifests a single probability peak which indicates a likely reliable photo-$z$ determination. The middle panel shows a doubly peaked EPDF where a redshift of $z \sim$ 0.2 and one of $\sim$2.8 are nearly equally probable, and the bottom panel shows an EPDF without clear peaks, both of which could indicate a possibly inaccurate photo-$z$ estimate. EPDFs are discussed in \S \ref{intro}.}
\label{epdf}
\end{center}
\end{figure}

The traditional spectroscopic method for determining redshifts is costly in terms of telescope time and therefore not practical for upcoming large-scale surveys. Redshift determinations for most galaxies in these surveys will instead rely on a more time efficient method of redshift estimation termed photometric redshift --- abbreviated ``photo-$z$'' --- see e.g. \citet{Salvato} for a recent review. While faster, this method is less accurate because it is based on a galaxy's measured brightness in a smaller number of wider wavelength bands of light compared with the traditional method. A crucial goal of the cosmological survey community is to develop methods of accurate photo-$z$ estimation with well understood statistical and systematic uncertainties \citep[e.g.][]{wp}. 

Photometric redshifts are typically calculated alongside estimates of their uncertainties. Unfortunately, however, for most photo-$z$ estimation techniques there are a small fraction of galaxies which have a true redshift which is more different from the estimated redshift than would be expected from the calculated uncertainty. These galaxies are often categorized into different degrees of ``outliers.'' Here we follow convention \citep[e.g.][]{Hildebrandt10} and define ``outliers'' as those galaxies where
\begin{eqnarray}
O: {{\vert z_{\rm phot}-z_{\rm spec} \vert} \over {1+z_{\rm spec}}} > 0.15,
\label{erroreq}
\end{eqnarray}
where $z_{\rm phot}$ and $z_{\rm spec}$ are the estimated photo-$z$ and actual (spectroscopically determined if available) redshift of the object. For the most inaccurate photo-$z$ estimations the term ``catastrophic outliers'' (hereafter COs) is often used. Although there is not a standard, universal definition of catastrophic outliers, we use a definition that is typical \citep[e.g.][]{BH10}:
\begin{eqnarray}
CO: {{\vert z_{\rm phot}-z_{\rm spec} \vert}} > 1.0.
\label{erroreqCO}
\end{eqnarray}
Those galaxies whose photo-$z$ estimates are sufficiently close to the actual redshift to not be outliers by the definition of equation \ref{erroreq} are termed ``non-outliers'' (NOs).  We note that because of the 1+$z_{\rm spec}$ in the denominator of equation \ref{erroreq} it is technically possible for a galaxy with a redshift of \textasciitilde 6 or greater to be defined as a CO but also an NO. However this work does not include any galaxies for which this is possible.

Because COs can have detrimental effects on the science goals of large scale surveys \citep[e.g.][]{Graham18,Hearin10, BH10}, mitigating them is therefore a crucial aim. One promising strategy could involve a method of identifying or `flagging' potential catastrophic outliers, so that these galaxies could be excluded or de-weighted in statistical analyses. 

One traditional class of photo-$z$ estimation techniques are the so-called ``empirical" or ``training set" methods, which work by developing a mapping from input parameters to redshift using a training set of galaxies whose redshifts are known, and then applying these mappings to an evaluation set \citep[e.g.][]{Hildebrandt10} for which the redshifts are to be determined. Empirical methods are often contrasted with ``template fitting'' methods which involve correlating the observed band photometry with model galaxy spectra and redshift. Empirical techniques encompass a variety of approaches including those involving machine learning. For this analysis we utilize SPIDERz (SuPort vector classification for IDEntifying redshifts), a custom support vector machine (SVM) type machine learning classification algorithm -- an example of an empirical method -- which is available to the community\footnote{available from http://spiderz.sourceforge.net with usage documentation provided there}. As is typical in empirical photo-$z$ techniques, SPIDERz takes as input galaxy photometric magnitudes or colors (and, optionally, any other parameters which may be relevant such as those quantifying shape information) and outputs an estimated redshift. As discussed in \citet[]{Jones17, Jones18}, SPIDERz has many customizable features, provides accurate redshift estimates on a variety of test data sets, and, crucially for this work, provides an effective redshift probability density function (EPDF) for each galaxy, which comes about naturally as a product of the SVM classification. The SVM works by comparing each possible redshift for a galaxy, divided into a user-inputted number of bins, and `voting' for the more likely option for each pair, which naturally provides an EPDF for the photometric redshift of each galaxy consisting of the total votes for each bin. In other words, the EPDF is an unknown monotonic function of a true probability density, and a relatively higher (lower) EPDF value always corresponds to a relatively higher (lower) probability. Bins that are unlikely to be accurate are paired against each other meaning that this method artificially inflates some bins with lower probabilities of being correct. Because of this it is not a true probability density function, but an EPDF -- however, the methods presented here will demonstrate that an EPDF can still be useful. Some example EPDFs for individual galaxies are shown in Figure \ref{epdf}. For a single-valued discrete photo-$z$ estimate, SPIDERz chooses the most probable bin of the EPDF. While in some cases the best strategy is to preserve the entire PDF and propagate that through analyses \citep[e.g.][]{ST13}, in other cases, such as publishing a catalog or because of data or processing capacity limitations in the case of millions of galaxies, a point redshift estimate may be necessitated. The strategies outlined in this work may be of use in the latter cases.

In a previous work \citep{Jones18} SPIDERz's naturally occurring EDPFs were noted as a method of potentially identifying and flagging potential COs, because galaxies with multiple widely separated peaks in redshift probability or without a clearly most probable redshift are more likely to be COs than those exhibiting a clear peak in probability, as can be visualized in Figure \ref{epdf}.  Preliminary results obtained there were promising, indicating that a relatively large fraction of COs ($\sim$50\%) could potentially be flagged while only flagging a relatively small fraction of the total NOs (\textless10\%).  In this work we will systematically explore the use of EPDFs as a method of flagging COs in photo-$z$ determination using a range of test data sets, and use this in conjunction with a modification of the redshift distribution of training sets in order to provide a realistic potential example `prescription' for achieving robust, accurate, and well understood photo-$z$ estimates in future large-scale surveys where the effect of COs --- especially at high redshifts --- is reduced. Although in a real survey scenario the redshifts of the evaluation sets would be unknown, here we use evaluation test sets with known redshifts in order to compare and test our results.

In \S \ref{tdata} we present the test data sets used throughout this work. In \S \ref{photozreal} we discuss some contemporary photo-$z$ challenges and consequently the various choices made for our testing, and what some photometric redshift studies have done that could be overly optimistic relative to future large-scale, several-band surveys with a large redshift range --- such as at least some of the galaxies from LSST.  In \S \ref{flagandopt} we describe a novel method for flagging potential outliers in estimations and our method for optimizing this. In \S \ref{mod} we explain our training set modification, used in order to produce fewer outlier galaxies as well as better flagging results using our algorithm. Finally, in \S \ref{pre} we present our example prescription for use in a realistic large-scale survey scenario. We emphasize that in \S \ref{photozreal}, \S \ref{flagandopt}, and \S \ref{mod} we utilize test data (described in \S \ref{tdata}) where all redshifts are known, in order to determine quantitatively the efficacy of the methods discussed.  In \S \ref{pre} by contrast, we consider the situation of an actual large-scale photometric survey in which the actual redshifts are not known.

\section{Test Data} \label{tdata}

\begin{figure} 
\begin{center}
\hspace*{-0.14in}
\includegraphics[width=3.00in]{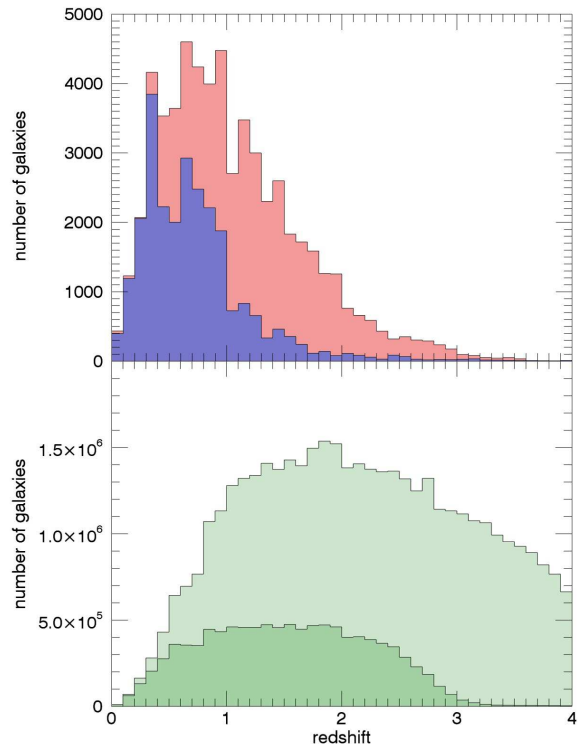}
\caption{The redshift distributions [$N(z)$] for the three test data sets used in this analysis. The top graph shows the distribution for the COSMOS-reliable-$z$ test data set (red) and the COSMOSxSpecs test data set (blue). The bottom graph shows the distribution for the BzK simulated test data set, where the lighter shade indicates the full data set and the darker shade shows the distribution after limiting all magnitudes at 28, as discussed in \S \ref{tdata}. }
\label{z_dist}
\end{center}
\end{figure}

In contrast to actual photo-z determinations in large-scale surveys where the desired redshifts are unknown a-priori, in order to explore photo-z methodologies the techniques are typically tested on galaxies where the actual redshifts are known, in order to explore the resulting accuracy.  In order to systematically explore using EPDFs as a method of flagging COs in photo-$z$ determination and the modification of training set distributions introduced above, we require test data sets with a large number of galaxies with photometric data as well as known redshifts over a wide range of redshifts and galaxy types.  These test sets could consist of either realistically simulated or real observed galaxy data.

One test data set of real galaxies comes from the COSMOS2015 photometric catalog \citep{Laigle16} with a separate redshift estimation collection \citep{Ilbert09}, which contains particularly reliable redshifts derived from a very large number of photometric bands --- such a large number over such a large wavelength range that it approaches what we will term a `quasi-spectroscopic redshift.' The COSMOS2015 catalog provides photometry for some galaxies in up to 31 optical, infrared, and UV bands in addition to redshifts calculated from this photometry, so to maximize the reliability of the known quasi-spectroscopic redshifts for our photo-$z$ test purposes, we restrict the use of galaxies to those meeting the following criteria: (i) non-error magnitude values for at least 30 bands of photometry, (ii) for which the ${\chi}^2$ for the \citet{Ilbert09} redshift estimate is $<$ 1, and (iii) for which the photo-$z$ value from the minimum ${\chi}^2$ estimate is less than 0.1 redshift away from the photo-$z$ value from the peak of the pdf. These galaxies can be considered to have highly reliable previous quasi-spectroscopic redshift estimates. Applying these criteria results in a data set of 58,622 galaxies. We then limited the redshifts to $z<4$ in order to prevent the occurrence of unoccupied bins, which resulted in a total of 58,619 galaxies.  We will refer to this set as `COSMOS-reliable-$z$.' For our test purposes, this test data set contains photometry for at least nine optical and infrared bands for each galaxy ($u$, $B$, $V$, $r$, $i$, $z$, $Y$, $J$, $K$), but for most purposes in this work we restrict photo-$z$ estimation to the use of only five optical bands ($u$, $B$, $r$, $i$, and $z$). 

\begin{table*}
\begin{center}
\scriptsize
\caption{Summary of test data set properties}
\label{tab0}
\begin{tabular}{cccc}
\hline
\hline
Name & Type & Number of Sources & $i$-band mag range (median) \\
\hline
`COSMOS-reliable-$z$' & `quasi-spectroscopic' redshifts + photometry & 58,619 & 27.11-19.00 (24.08) \\
`COSMOSxSpecs' & spectroscopic redshifts + photometry & 25,831 & 26.67-18.00 (22.41) \\
`BzK' & simulated redshifts \& photometry & 10 million & 28.18-9.75 (25.96) \\
\hline
\end{tabular}
\tablecomments{\footnotesize{Summary of test data sets used in this work, as discussed in \S \ref{tdata}.  Redshift distributions are shown in Figure \ref{z_dist}.   As discussed in the text we use `quasi-spectroscopic' here to refer to an especially reliable photometric redshift derived from 30 wavebands extending from the infrared to the ultraviolet which we take as equivalent to a spectroscopic redshift. In addition the BzK data set has been limited to objects brighter than an $i$-band magnitude of 28 and has added noise.}}
\end{center}
\end{table*}

In order to form another test data set of real galaxies with a somewhat different redshift distribution, we utilize a combination of several spectroscopic redshift collections \citep{Hasinger18, Salvato18, Momcheva16} with the COSMOS catalog of magnitudes \citep{Laigle16}, which together forms a data set complete with 9 photometric magnitude bands (of which we generally utilize five for photo-$z$ estimation --- $u$, $g$, $r$, $i$, and $z$). For spectroscopic redshifts we use the reported "best available" value and eliminate galaxies which are flagged as having their redshift determined from photometry or as being stars. Limiting the redshifts to $z <$ 4, this results in a test data set of 25,831 galaxies. We will refer to this set as `COSMOSxSpecs.'

For a still different redshift and galaxy type distribution, we utilize simulated data from the BzK Deep lightcone mock catalog, with over 41 million simulated galaxies \citep{bzk}, as discussed in \S \ref{photozreal}. We utilize the simulated photometric observations in the $u$, $g$, $r$, $i$, and $z$ bands for photometric redshift estimation tests. We note that this photometry is available in bandpasses that are quite similar to the bandpasses in the real galaxy data sets discussed above.  This data in its raw form resulted in almost no CO photo-$z$ estimates, which is itself an indication of the potential perils of photo-$z$ studies using simulated data. In order to make the data set more usable for this investigation, which requires some CO estimates, it was necessary to add simulated noise. We altered each magnitude value by a random positive or negative amount, distributed evenly, between zero and 15\%. We also removed all galaxies dimmer than magnitude 28 in any band in order to better reflect real large-scale survey data sets, resulting in a test data set of over 10 million galaxies.

The redshift distributions of the test data sets are shown in Figure \ref{z_dist} and the parameters are summarized in Table \ref{tab0}.

\section{Some Contemporary Photo-$z$ Challenges} \label{photozreal}

Some studies of photometric redshift estimation techniques with test data sets (referring here to studies of the techniques themselves and not the photo-$z$ catalogs of large-scale surveys) have been non-representative to varying degrees of certain challenges that will arise from upcoming large-scale surveys with a limited number of photometric bands and for which redshift estimates will be needed for galaxies spanning a large range of redshifts. It will be vital for e.g. weak lensing science goals to perform photo-$z$ estimations using just the available measured magnitudes in several optical bands and to accurately estimate redshifts for galaxies extending to $z \sim 2$ and beyond. Importantly, we note that galaxies with redshifts higher than this, even if not directly the most relevant to weak lensing studies, are important because they can be mis-classified as lower redshift galaxies if COs, thus affecting other science goals. Because of this, we emphasize the importance of galaxies extending to redshifts $z \sim 4$. In addition, in realistic scenarios the training set will be very small compared to the potential evaluation set, so we utilize a training set that is significantly smaller in number than the evaluation set. Here we aim to motivate an example prescription which can be applied to photo-$z$ estimations in future large-scale surveys in order to mitigate the effects of CO estimates.
 
\begin{figure} 
\begin{center}
\hspace*{-0.14in}
\includegraphics[width=3.00in]{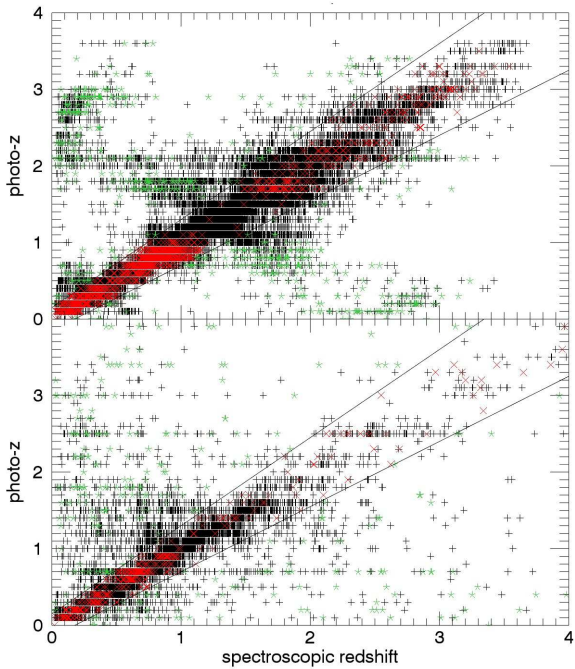}
\caption{Comparison of the estimated vs. actual redshift for a nine-band (top) and five-band (bottom) evaluation using COSMOS-reliable-$z$ test data with 2,000 training set galaxies in each case. The five band case uses $u$, $B$, $r$, $i$, and $z$ photometric bands, while the nine band case uses $u$, $B$, $V$, $r$, $i$, $z$, $Y$, $J$, $K$. This figure highlights the degradation seen when estimating based on five optical photometric bands as opposed to nine bands which extend into the near infrared, as discussed in \S \ref{photozreal}. Those points outside of the inner diagonal (lighter) lines are outlier photo-$z$ estimates as defined by equation \ref{erroreq}, while those outside of the outer (heavier) lines are CO photo-z estimates as defined by equation \ref{erroreqCO}. The nine-band analysis results in 4.05\% outliers with 1.06\% COs, while for the five-band case these percentages are 9.94\% and 2.80\%, respectively. }
\label{9v5}
\end{center}
\end{figure}

It is well established that the presence or absence of additional infrared bands has a significant effect on photo-$z$ estimation accuracy. As an example of this, Figure \ref{9v5} highlights a case of the degradation baseline photo-z estimation when reducing from nine bands extending into the infrared to five optical-only bands. Some large scale surveys do not include infrared bands.  LSST does not in its default survey mode, although there will be some overlap with Euclid \citep[e.g.][]{Rhodes19}.   In this work we focus on the use of five optical-only bands ($u$, $g$ or $B$ as available, $r$, $i$ and $z$).  While this may seem overly pessimistic for LSST in particular which will feature six photometric bands (roughly the five preceding and a near-infrared $y$ band in addition), and will have some infrared overlap as mentioned above, in reality many galaxies will not have overlap and will not have photometry in all six bands and/or will report large photometric errors in one or more bands \citep[e.g.][]{Soo18}. Therefore we view five photometric bands as neither overly optimistic nor overly pessimistic for many galaxies.
 
Another potential scenario for large-survey photo-$z$ estimation --- unless measures are taken specifically to avoid it --- is one in which the set of galaxies with spectroscopic redshifts available --- i.e. those which can comprise a training set in empirical photo-$z$ estimation --- is not representative of the galaxies for which photo-$z$ estimations are desired. To test this, we use a training set comprised of galaxies from one test data set and then evaluate on a subset of galaxies from a different test data set. Then, in an attempt to improve the estimation in a way that would be possible if the redshifts of the evaluation set were truly unknown, we alter the training set to match the distribution of the $i$-band magnitudes by bins of 0.1 of the evaluation set. Despite the matching $i$-band distributions the photo-$z$ accuracy is poor in this situation, as shown in Table \ref{tab1}. These `mismatched` training and evaluation sets perform very poorly even when using the very complete, deep, simulated BzK catalog as a training set, despite BzK data having a redshift distribution which covers a much larger range compared to the two real-galaxy COSMOS test data sets, and thus potentially containing more combinations of galaxy type and redshift than the COSMOS sets. This indicates that training and evaluation set mismatches in galaxy types and other characteristics creates problems for empirical photo-$z$ approaches. Because of this, we believe that it is essential that for future large scale surveys, the photo-$z$ training set be formed from a subset that is as representative as possible of the overall galaxy sample for which photo-$z$s are desired. We return to this issue in \S \ref{pre}.

\section{Flagging and Flagging Optimization} \label{flagandopt}

Our method for using EPDFs to flag potential COs uses the heights and locations of peaks within the EPDFs, some examples of which are shown in Figure \ref{epdf}. The flagging procedure first identifies the maximum peak of a galaxy's EPDF, which corresponds to the most likely redshift and therefore the discrete single-valued photo-$z$ estimate. If a second peak in redshift is above a predetermined minimum fraction of the height of the primary peak ($p_{\rm f,min}$) and farther than a predetermined minimum distance away in redshift from the maximum peak ($\Delta z_{\rm peak,min}$), the galaxy is flagged as a potential outlier. This idea was explored preliminarily in a previous work \citep{Jones18} and was found to be a promising strategy. Here, rather than attempting to find one optimal combination of these parameters for all data configurations, we instead optimize these parameters for each data configuration individually. The ultimate goal of the optimization of these parameters is to flag more COs while simultaneously flagging fewer NOs. However, simply increasing the ratio of flagged COs to flagged NOs generally results in an unacceptable number of flagged NOs in high redshift ($z>1.5$) bins. Due to the importance of galaxies of high redshift in many science goals, we wished to avoid this. Keeping this in mind, we utilize the following penalty function which is the ratio of the number of flagged NOs to the total number of NOs greater than spectroscopic redshift $z = 1.5$:

\begin{figure} 
\begin{center}
\hspace*{-0.14in}
\includegraphics[width=3.25in]{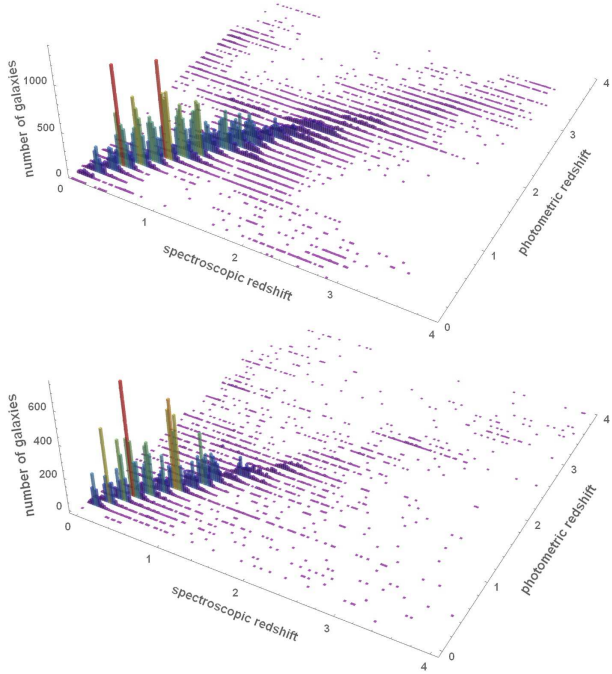}
\caption{Estimated photo-$z$ vs. actual redshift as evaluated with SPIDERz with five photometric bands using COSMOS-reliable-$z$ (top) and COSMOSxSpecs (bottom) test data, with training sets as modified using the methods discussed in \S \ref{mod}. Those points outside of the inner diagonal (lighter) lines are outlier photo-$z$ estimates as defined by equation \ref{erroreq}, while those outside of the (heavier) outer lines are CO photo-z estimates as defined by equation \ref{erroreqCO}. Shown in red are erroneously flagged NOs and shown in green are correctly flagged Os using the optimal flagging parameters found with the flagging method discussed in \S \ref{flagandopt}. Given the density of galaxies in some of the regions of these plots, we also show the same results with a density visualization in Figure \ref{flag_3d}.}
\label{flag_2d}
\end{center}
\end{figure}

\begin{figure} 
\begin{center}
\hspace*{-0.14in}
\includegraphics[width=3.25in]{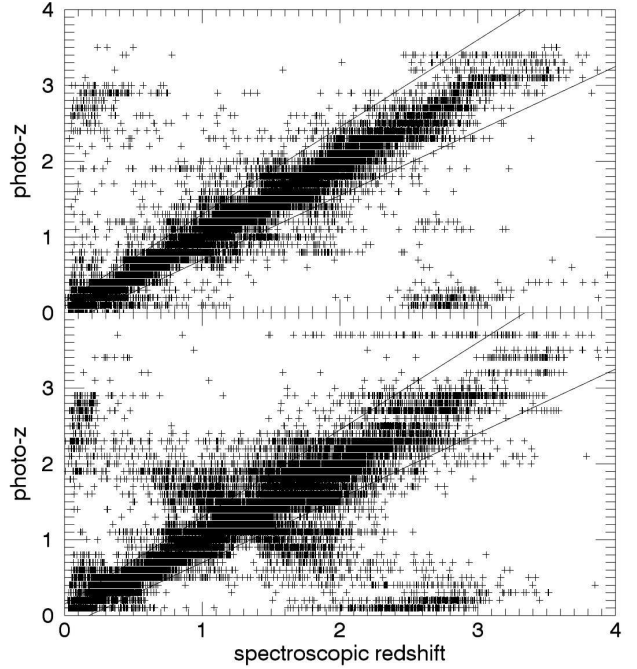}
\caption{Comparison of the estimated vs. actual redshift as in Figure \ref{flag_2d} but with a 3-D visualization where the height of a bar indicates the number of galaxies in that bin. The spectroscopic redshifts and photo-$z$ estimates are visualized here binned in intervals of 0.25 in redshift.}
\label{flag_3d}
\end{center}
\end{figure}

\begin{figure} 
\begin{center}
\hspace*{-0.14in}
\includegraphics[width=3.00in]{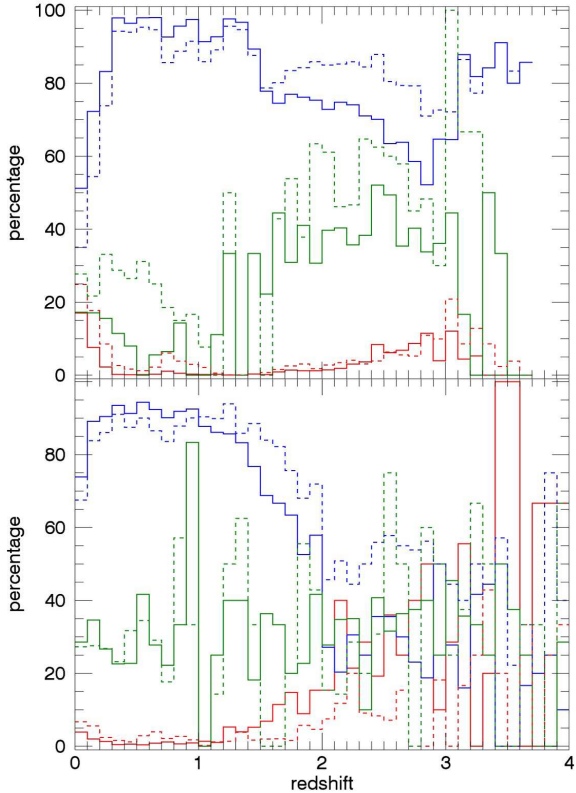}
\caption{Several flagging reliability metrics for representative determinations using the COSMOS-reliable-$z$ (top) and COSMOSxSpecs (bottom) test data sets. We show results both for an 8,000 training set ``baseline case" (solid lines) and a modified randomly undersampled (discussed in \S \ref{mod}) training set case (dashed lines). We show the percentage of flagged COs of the total number of COs in each redshift bin (green), the percentage of flagged NOs of the total number of NOs in each redshift bin (red), and the percentage of non-flagged NOs of the total number of galaxies in each redshift bin (blue). While the results of flagging on the base case are already significant, these graphs and metrics also show a clear improvement with utilization of the modified training set over the baseline 8,000 galaxy training set at higher redshifts, without significant degradation at lower redshifts.  For both test data sets the modified training set results in a significant increase in the percentage of non-flagged NOs at higher redshifts, as well as an increase in the percentage of flagged COs at high redshifts in the COSMOS-reliable-$z$ test data set and a decrease in the percentage of flagged NOs at high redshifts in the COSMOSxSpecs test data set.}
\label{flaggingperc}
\end{center}
\end{figure}

\begin{eqnarray}\label{pf}
f_{\rm P} = \frac{NO_{\rm flagged} > 1.5}{NO_{\rm total} > 1.5}.
\end{eqnarray}
This penalty function is then applied to a `goodness function' which is simply the ratio of the total percentage of flagged COs to the penalty function:

\begin{eqnarray} \label{goodness}
f_{\rm G} = \frac{\%CO^{\rm flagged}}{f_{\rm P}}.
\end{eqnarray}
Each combination of parameters $p_{\rm f,min}$ and $\Delta z_{\rm peak,min}$ is then checked via a grid search to maximize the goodness function, with the addition of a minimum cutoff of 30\% flagged COs. Some visualizations of the effectiveness of flagging in this way can be seen in Figures \ref{flag_2d} and \ref{flag_3d}. We present quantitative metrics describing the efficacy of flagging in this way in Tables \ref{tab1} and \ref{tab2}.  Additionally, some visualizations of these metrics as a function of redshift are shown in Figure \ref{flaggingperc}, noting though that these figures also present how flagging efficacy is altered by the training set modification discussed later in \S \ref{mod}.  We see that this method of flagging according to optimized $p_{\rm f,min}$ and $\Delta z_{\rm peak,min}$ values can effectively flag a large portion of COs while simultaneously flagging a much smaller percentage of NOs, and relatively small percentages of NOs even in high redshift bins, depending on the test data set. We note that in this analysis the optimal flagging parameters were determined with knowledge of the actual redshifts for all galaxies in our estimation set. In a real large-scale survey photometric redshift estimation campaign, the flagging parameter optimization would have to be carried out on a subset of galaxies for which actual redshifts are available, as discussed in \S \ref{pre}.

\section{Training Set Modification} \label{mod}

\begin{figure} 
\begin{center}
\hspace*{-0.14in}
\includegraphics[width=3.00in]{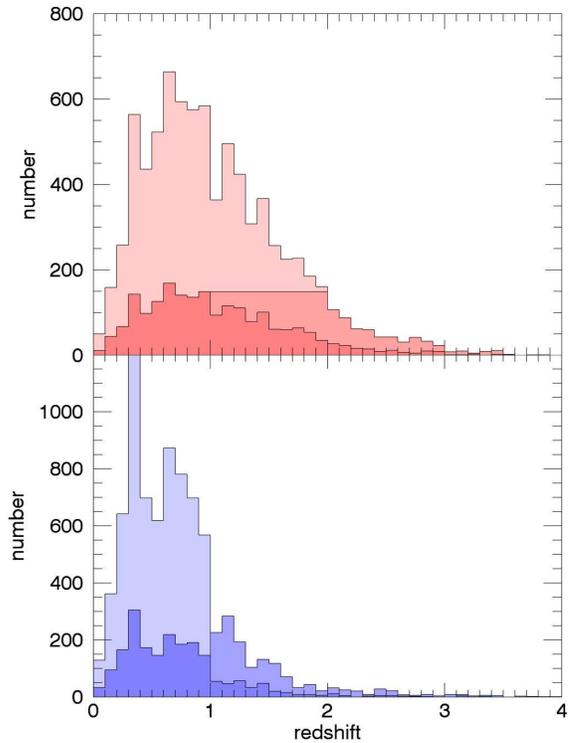}
\caption{Several representative example redshift distributions for training sets for the COSMOS-reliable-$z$ (top) and COSMOSxSpecs (bottom) test data sets. Training set modifications are discussed in \S \ref{mod}. The lightest shade shows the redshift distribution for an 8,000 galaxy `baseline case' training set while the darker shade shows the modified randomly undersampled training set, and the darkest shade shows the sample 2,000 galaxy set which was used in order to form the modified sets.}
\label{rsmods}
\end{center}
\end{figure}

Here we explore modifying the redshift distribution of training sets in order to reduce the number of COs and to increase the efficacy of flagging. An example of one modification for each data set can be seen in Figure \ref{rsmods}. Some methods of training set modification such as deweighting have been explored in the literature \citep[e.g.][]{AS,Hat}. In this work the modification of training sets involves creating a distribution of galaxies in redshift across the entire range of redshifts that is less heavily skewed toward lower redshifts by removing some lower redshift galaxies. In other words, we seek to increase the \textit{fraction} of galaxies in the training set which are in the highest (usually less populated) redshift bins. This is a particular implementation of what can be referred to in data analysis and machine learning as ``random undersampling.'' In the context of machine learning, random undersampling involves reducing classes which are more populated  in the training set in order to make less populated classes \textit{proportionally} more prominent, so we will label the resulting modified training sets `randomly undersampled.'

\begin{figure} 
\begin{center}
\hspace*{-0.14in}
\includegraphics[width=3.00in]{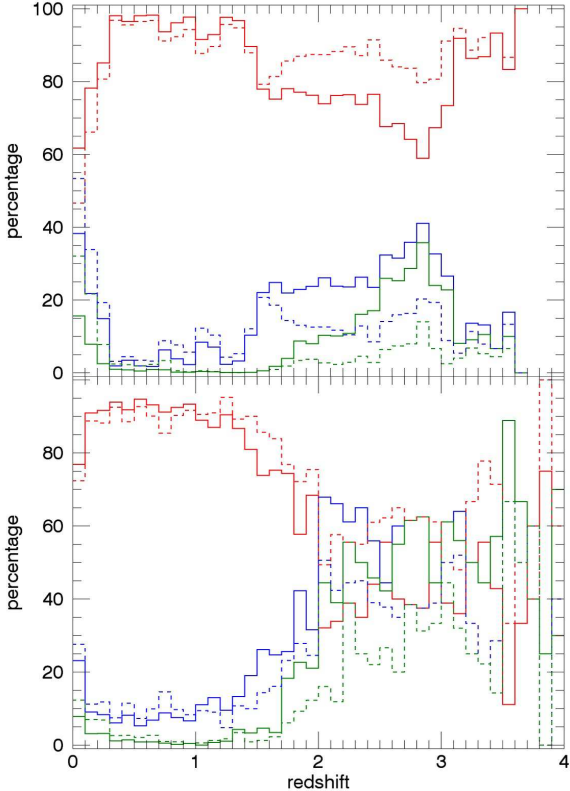}
\caption{Photo-$z$ reliability metrics for representative determinations using the COSMOS-reliable-$z$ (top) and COSMOSxSpecs (bottom) test data sets for an 8,000 training set `baseline case' (solid lines) and a modified randomly undersampled training set case (dashed lines). These training set modifications are discussed in \S \ref{mod}. We show results for the percentage of NOs of the total number of galaxies in each redshift bin (red), and the percentage of outliers (blue) and of COs (green) of the total number of galaxies in each redshift bin. These metrics show a clear improvement with the use of the modified training set over the baseline 8,000 galaxy training set: the modified training sets result in a significant increase in the percentage of NOs, and a decrease in Os and COs for both test data sets at higher redshifts, without significant degradation at lower redshifts.}
\label{percplots}
\end{center}
\end{figure}

With an eye toward formulating a useful example prescription for large-scale surveys as discussed in \S \ref{pre} we implement the modification in the following manner: we first randomly select a separate 8,000 and 10,000 galaxies from the full test data set, the former for a training set and the latter for the evaluation set. 2,000 galaxies are also randomly selected from the 8,000 galaxy subset to be used as a fiducial sample only for its redshift distribution. Because of the different redshift distributions of the COSMOS-reliable-$z$ and COSMOSxSpecs data sets, the randomly undersampled training set is then created slightly differently in the two cases. For the COSMOS-reliable-$z$ test data set, the distribution of the 2,000 galaxy set is used from $z =$ 0 to $z =$ 1, then the number of galaxies in the $z =$ 0.9 - 1.0 bin is used as the value for each successive bin until a lower number of galaxies in a bin is found, usually around $z =$ 2. The 8,000 galaxy distribution is used for all redshift bins above this. In the case of the COSMOSxSpecs test data set, the 2,000 galaxy training set distribution is used from $z =$ 0 to $z =$ 1 and the 8,000 galaxy training set distribution is used for the remaining bins. The appropriate number of galaxies are randomly selected for each bin from the 8,000 galaxy training set in order to fit the modified distribution as described. These modifications are visualized in Figure \ref{rsmods}. 

These modifications result in improvements in the overall reliability at higher redshifts (visualized in Figures \ref{flaggingperc} and \ref{percplots}) in the form of significant increases in the fraction of galaxies at high redshift which are non-flagged NOs --- i.e. those galaxies whose redshift estimations are correct and are not flagged and thus presumably ideal for science analyses, as well as significant decreases in the fraction of galaxies at high redshift which are non-flagged COs --- i.e. those galaxies which are incorrectly estimated and are not flagged. Several metrics relevant to both of these categories of improvements are also shown in Table \ref{tab2}. For both test data sets this is achieved in significant part by reducing the percentage of outliers and COs at high redshift dramatically, while in the case of the COSMOSxSpecs test data set the percentage of flagged NOs at high redshifts is also decreased dramatically, and in the case of the COSMOS reliable-$z$ test data set the percentage of flagged COs at high redshifts is increased significantly. In the case of COSMOS reliable-$z$ the percentage of flagged NOs at high redshift increases, but not by enough to counteract the effect of the dramatic increase in the number of NOs at high redshift (corresponding to the reduction in outliers), and thus the modification still results in a significant increase in the fraction of non-flagged NO high redshift galaxies. In the case of COSMOSxSpecs the percentage of flagged COs at high redshift decreases, but ---  again --- not by enough to counteract the effect of the dramatic decrease in the number of COs at high redshift, and thus the modification results in a significant decrease in the fraction of non-flagged CO high redshift galaxies.

%[[[reductions in the numbers of flagged NOs as well as increases in the percentages of flagged COs. Improvements in the general photo-$z$ reliability are visualized in Figure \ref{percplots}. We see a significant increase in the numbers of NOs in the range $z >$ 1 as well as decrease the number of COs in the range $z >$ 1.  As shown in Figure \ref{flaggingperc}, although in the case of the COSMOS-reliable-$z$ test data set the percentage of flagged NOs resulting from the training set modification does not decrease like in the case of the COSMOSxSpecs test data set but rather remains very similar, it is important to note that the percentage of non-flagged NOs above redshift $z =$ 1 of the total number of galaxies in each bin is increasing. This is because the modification causes more NOs at higher redshifts, so although in the case of COSMOS reliable-$z$, the percentage of flagged NOs remains similar, the \textit{number} of non-flagged NOs is increased. ]]]

It is important to note that these improvements do not come about simply by virtue of having a higher {\it number} of high redshift galaxies, but by having a higher {\it fraction} of high redshift galaxies. This modification does not involve increasing the number of high redshift galaxies compared to the baseline training set distribution, but rather removes some lower redshift galaxies as seen in Figure \ref{rsmods} resulting in a training set that is actually smaller in number (although to compose this training set an equal number of galaxies is needed to begin with, as discussed in \S \ref{pre}). In this analysis the training set modification is not necessarily completely optimized for each case; we present this as a relatively straightforward and achievable modification which can significantly improve CO identification as well as baseline photo-$z$ estimates, and believe that any further optimization would be subject to a data set's particular composition and therefore not generalizable.  However, we see that even not fully optimized random undersampling of a training set results in improvements in both baseline photo-$z$ performance and flagging, and furthermore the specific undersampling for any given data set could in principle be optimized as long as there were a sufficient number of galaxies with known redshifts (likely around 10,000 as discussed below) in order to optimize the flagging parameters.  

\section{Prescription} \label{pre}

\begin{figure} 
\begin{center}
\hspace*{-0.14in}
\includegraphics[width=3.00in]{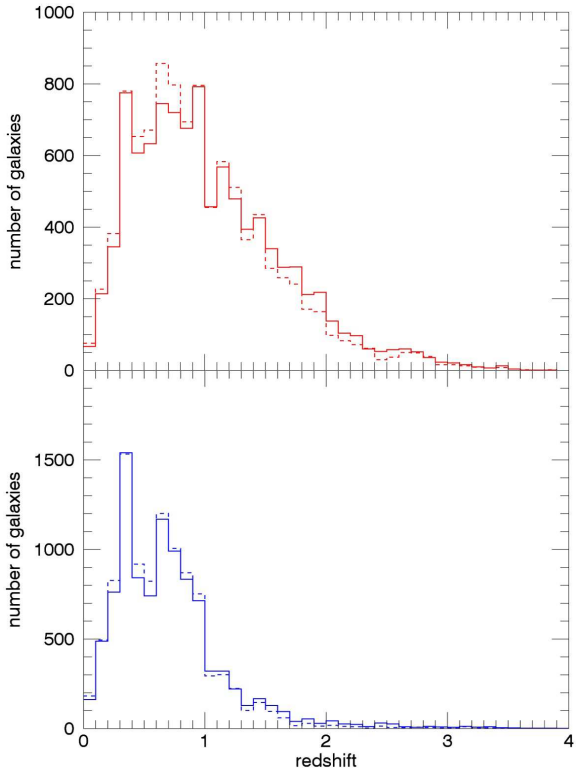}
\caption{Two representative redshift distributions for 10,000 galaxies randomly selected from COSMOS-reliable-$z$ (top) and COSMOSxSpecs (bottom). Solid lines represent the redshift distribution for 10,000 galaxies randomly selected from their full respective data sets while dotted lines show the distribution for 10,000 galaxies which were selected by first randomly selecting 11,111 galaxies and then removing the dimmest 10\% in $i$-band. The differences in these distributions highlight the differences that arise in a situation where some (in this case 10\%) of the dimmest galaxies' spectroscopic redshifts cannot be determined for use in a training set compared to one where no galaxies are lost, as discussed in \S \ref{pre}.}
\label{maglims}
\end{center}
\end{figure}

\begin{figure} 
\begin{center}
\hspace*{-0.14in}
\includegraphics[width=3.00in]{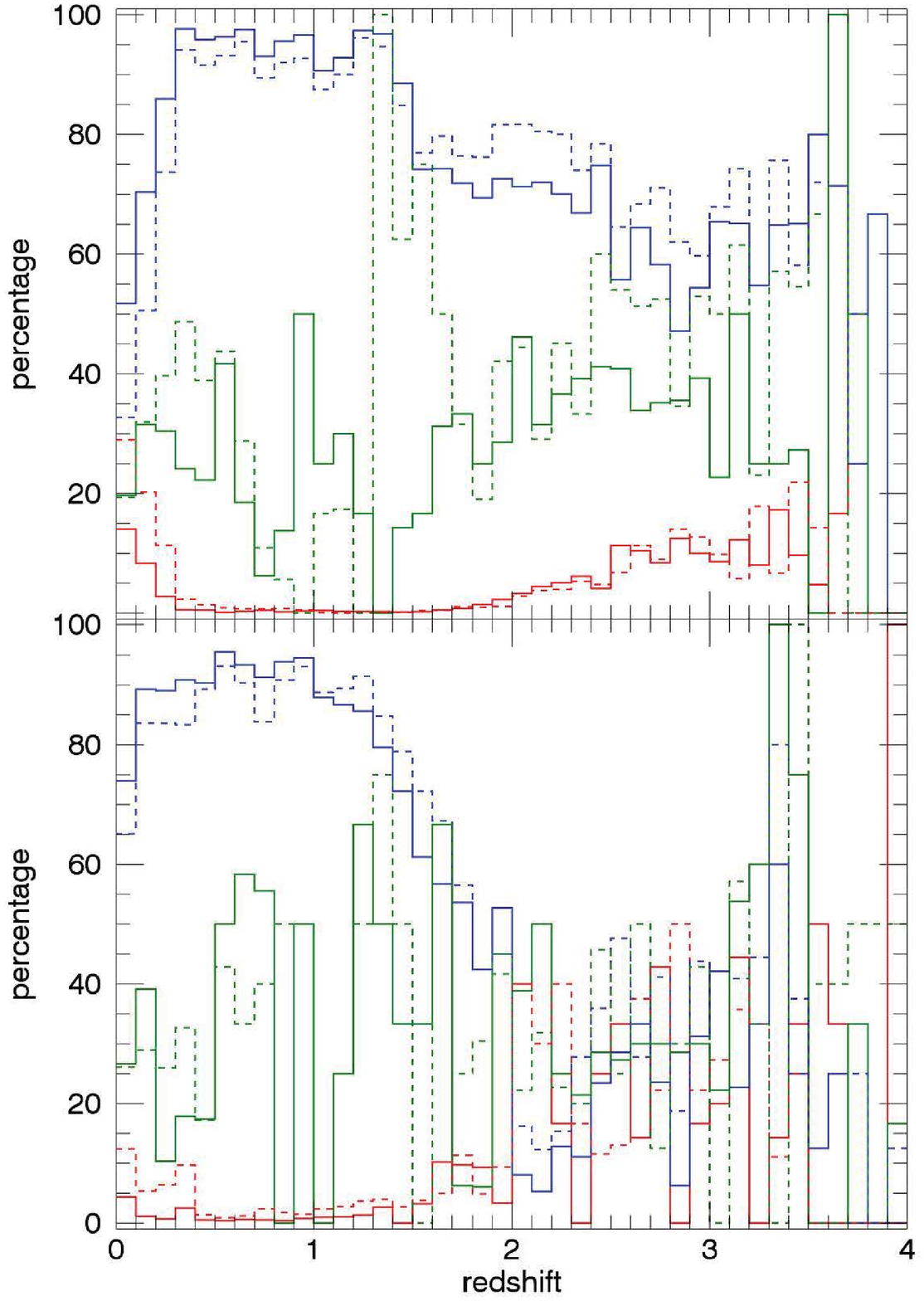}
\caption{The same flagging reliability metrics as Figure \ref{flaggingperc} but using the {\it spectro-magnitude-limited} COSMOS-reliable-$z$ (top) and COSMOSxSpecs (bottom) test data sets for an 8,000 training set `baseline case' (solid lines) and a modified randomly undersampled (discussed in \S \ref{mod}) training set case (dashed lines). Magnitude limitation is discussed in \S \ref{pre}. As with Figure \ref{flaggingperc} we show results for the percentage of flagged COs of the total number of COs in each redshift bin (green), the percentage of flagged NOs of the total number of NOs in each redshift bin (red), and the percentage of non-flagged NOs of the total number of galaxies in each redshift bin (blue).}
\label{maglimflaggingperc}
\end{center}
\end{figure}

\begin{figure} 
\begin{center}
\hspace*{-0.14in}
\includegraphics[width=3.00in]{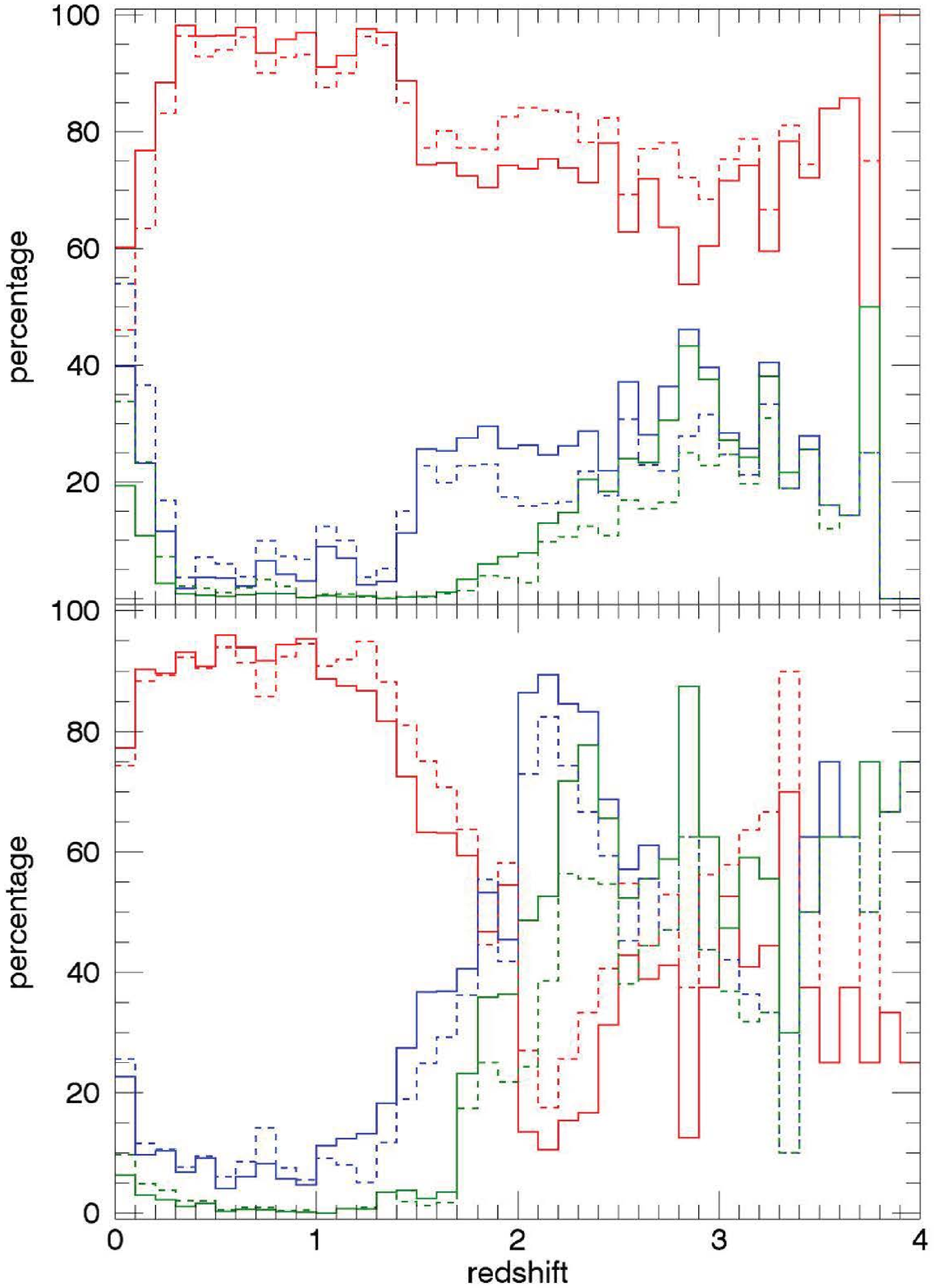}
\caption{The same photo-$z$ reliability metrics as Figure \ref{percplots} but using the {\it spectro-magnitude-limited} COSMOS-reliable-$z$ (top) and COSMOSxSpecs (bottom) test data sets for an 8,000 training set ``baseline case" (solid lines) and a modified randomly undersampled (discussed in \S \ref{mod}) training set case (dashed lines). Magnitude limitation is discussed in \S \ref{pre}. As with Figure \ref{percplots} we show results for the percentage of NOs of the total number of galaxies in each redshift bin (red), and the percentage of outliers (blue) and of COs (green) of the total number of galaxies in each redshift bin.}
\label{maglimpercplots}
\end{center}
\end{figure}

In section \S \ref{photozreal} we demonstrated that empirical photo-$z$ results are poor when a training set is not drawn from the same survey data as the evaluation set, while in \S \ref{flagandopt} and \S \ref{mod} we showed the possible utility of flagging potential COs and modifying training set redshift distributions when those training sets are drawn from the same survey as the evaluation set. Based on these conclusions and the results discussed in those sections and as seen e.g. in Table \ref{tab1} and \ref{tab2}, here we present an example prescription for achieving useful photo-$z$ estimation with reduced high redshift COs, efficiently flagged remaining COs, and increased high redshift non-flagged NOs in one realistic photometric estimation scenario for a hypothetical large-scale survey with a limited number of photometric bands, starting from a large set (potentially millions) of observed galaxies for which redshifts are unknown and desired to be estimated:
\begin{enumerate}
\item Obtain spectroscopic redshifts for a random subset of 18,000 galaxies.  This is the number necessary to achieve a 10,000 galaxy testing set plus a separate undersampled training set of sufficient size.  
\item Set aside a randomly selected 10,000 of these galaxies, to be used as an evaluation testing set for flagging parameter determination.  As discussed below, we determined that 10,000 is the minimum number of galaxies for randomly chosen evaluation testing sets that is likely to result in consistent optimized flagging parameters over multiple determinations.  
\item Randomly undersample the remaining 8,000 galaxies by removing some lower redshift galaxies, as described in \S \ref{mod}, for use as the training set.
\item Train on the training set and evaluate on the 10,000 galaxy subset selected earlier to find the optimal flagging parameters which maximize the goodness function as discussed in \S \ref{flagandopt}.
\item Estimate the redshifts of the millions of galaxies for which photometric redshifts are desired using the modified training set and flag potential COs using the found optimal parameters from step 4.
\end{enumerate}

This example prescription is operable in principle for any campaign in which photo-$z$s are determined with any empirical method which provides probability distribution over redshift information for each galaxy. 

Steps 3 and 5 are necessary because the optimization routine works by using known redshifts in order to calculate the percentages of flagged NOs and COs. Because in real situations redshifts are not known for the evaluation sets, it is necessary to create a smaller evaluation set with spectroscopic redshifts in order to optimize the parameters for the full evaluation. In the case of the COSMOS-reliable-z data set, we find that 10,000 galaxies is the smallest number that produces consistent results in terms of finding the same optimized flagging parameters $p_{\rm f,min}$ and $\Delta z_{\rm peak,min}$ as the full evaluation set, while in the case of the COSMOSxSpecs test data set we find that even 10,000 galaxy evaluation sets do not find consistent optimal parameter values across multiple randomizations. We have identified the cause of this as being certain galaxies which have a difference in a very small number (in some cases 1) of `votes' in certain bins of their EPDFs. This is ultimately based on the difference in scaling of the magnitudes that occurs when galaxies which are extremal in magnitude do or do not make it into the randomly selected evaluation set due to the small number of galaxies in high redshift bins in the COSMOSxSpecs test data set. We note also that the result of 10,000 galaxies being the likely number of evaluation set objects necessary to achieve consistent optimal parameter values, and thus the need for 18,000 spectra overall, is for a photo-$z$ parameter space of five photometric bands and may be different to some degree for the case of more or fewer bands.

\begin{table*}
\begin{center}
\scriptsize
\caption{Photo-z performance and flagging metrics for various test data set combinations}
\label{tab1}
\begin{tabular}{rrrrrrrrrrrr}
\hline
\hline
\textbf{9 BAND:}& & & & & & & & & & & \\
\hline
Test Data& &$O$\%&$CO$\%&$O$\%$>$1.5&$CO_{\rm f}$\%&$NO_{\rm f}\%$&$NO_{\rm f}\%$$>$1.5&$NO_{\rm f}\%$$>$2.0&$\%NO_{\rm nf}$$>$1.5&$p_{\rm f,min}$&$\Delta z_{\rm peak,min}$\\
\hline
xSpecs& &7.77&1.65&34.23&35.85&1.24&6.83&13.17&61.28&0.91&1.10\\
reliable-$z$& &4.05&1.06&6.85&45.77&0.94&1.91&5.20&91.37&0.94&1.20\\
\hline
\hline
\textbf{5 BAND:}& & & & & & & & & & & \\
\hline
\textbf{Standard Runs}& & & & & & & & & & & \\
Test Data& & & & & & & & & & & \\
\hline
xSpecs& &11.69&3.50&41.44&30.01&1.57&13.06&18.75&50.91&0.95&0.90\\
reliable-z& &9.94&2.80&21.76&35.23&1.61&4.15&9.92&74.99&0.94&1.20\\
BzK& &7.34&0.73&8.41&38.18&0.47&0.57&0.64&91.07&0.94&1.20\\
\hline
\textbf{Mismatched}& & & & & & & & & & & \\
Train. Set&Eval. Set& & & & & & & & & & \\
\hline
xSpecs&reliable-$z$&15.13&5.94&72.32&36.71&6.60&31.68&52.94&18.91&0.94&1.10\\
xSpecs&BzK&15.18&4.66&39.22&48.84&5.12&10.84&17.27&54.19&0.95&1.20\\
reliable-$z$&xSpecs&35.57&10.32&23.53&31.14&7.47&5.67&11.05&72.13&0.90&1.10\\
reliable-$z$&BzK&35.90&13.33&43.94&38.35&7.21&8.95&12.89&51.05&0.91&1.10\\
BzK&xSpecs& 19.21&5.38&58.24&32.52&1.73&9.68&32.74&37.72&0.90&1.20\\
BzK&reliable-$z$&21.11&5.54&75.72&30.18&0.93&24.48&74.36&18.33&0.80&1.10\\
\hline
\end{tabular}
\end{center}
\tablecomments{\footnotesize{
Shown are various metrics for determinations run using 9 and 5 bands of photometry. ``Standard Runs'' indicate determinations run using an unmodified 2,000 galaxy training set. Tallied for each determination are the percentage of galaxies which are outliers ($O\%$), the percentage of galaxies which are catastrophic outliers ($CO\%$), the percentage of galaxies above $z =$ 1.5 which are outliers ($O\%>1.5$), the percentage of catastrophic outliers which are flagged ($CO_{\rm f}\%$), the percentage of non-outliers which are flagged ($NO_{\rm f}\%$), the percentage of non-outliers above $z =$ 1.5 which are flagged ($NO_{\rm f}\%>1.5$), the percentage of non-outliers above $z =$ 2.0 which are flagged ($NO_{\rm f}\%>2.0$), the percentage of galaxies above $z =$ 1.5 which are non-flagged non-outliers ($\%NO_{\rm nf}>1.5$), and the optimal parameters used for flagging ($p_{\rm f,min}$ and $\Delta z_{\rm peak,min}$) determined using the methods discussed in \S \ref{flagandopt}. Relevant to \S \ref{photozreal} we show results for determinations with 9 photometric bands, with the simulated BzK test data set, and ``mismatched'' runs which use training sets drawn from different test data sets than the paired evaluation set.}}
\end{table*}
 
\begin{table*}
\scriptsize
\caption{Photo-z performance and flagging metrics for several training set modifications}
\label{tab2}
\begin{center}
\begin{tabular}{rrrrrrrrrrrrrrr}
\hline
\hline
\textbf{Train. Mods.}& & & &$O$\%&$CO$\%& &$CO_{\rm f}$\%&\%$CO_{\rm nf}$& &$NO_{\rm f}\%$&$NO_{\rm f}\%$&$\%NO_{\rm nf}$& & \\
Test Data&Mod.&$O$\%&$CO$\%&$>$1.5&$>$1.5&$CO_{\rm f}$\%&$>$1.5&$>$1.5&$NO_{\rm f}\%$&$>$1.5&$>$2.0&$>$1.5&$p_{\rm f,min}$&$\Delta z_{\rm peak,min}$\\
\hline
xSpecs&Baseline&10.31&2.93&41.86&26.74&37.16&35.45&17.26&1.57&15.85&43.39&48.93&0.93&1.00\\
xSpecs&Undersampled&11.31&3.35&28.98&13.33&30.15&33.56&8.86&2.60&5.92&12.22&66.82&0.93&1.00\\
reliable-$z$&Baseline&9.36&2.56&24.47&8.82&30.99&38.43&5.43&0.90&2.15&4.36&73.91&0.97&0.90\\
reliable-$z$&Undersampled&9.60&2.55&15.03&2.94&33.93&48.39&1.52&2.51&2.88&6.38&82.52&0.91&1.20\\
\hline
\hline
\textbf{$i$-band Lim.}& & & & & & & & & & & & & & \\
Test Data&Mod.& & & & & & & & & & & & & \\
\hline
xSpecs&Baseline&11.26&3.12&53.85&32.23&30.11&30.68&22.34&1.39&12.10&28.35&40.57&0.94&0.90\\
xSpecs&Undersampled&11.55&3.05&44.78&22.99&32.69&33.47&15.29&4.04&11.11&24.31&49.08&0.91&0.90\\
reliable-$z$&Baseline&10.18&2.78&27.61&9.32&31.62&34.93&6.06&1.09&3.01&6.62&70.22&0.95&1.10\\
reliable-$z$&Undersampled&11.42&3.18&21.04&5.81&33.63&42.98&3.31&1.88&2.82&6.26&76.73&0.92&1.20\\
\hline
\end{tabular}
\end{center}
\tablecomments{\footnotesize{
Tallied for each determination are the percentage of galaxies which are outliers ($O\%$), the percentage of galaxies which are catastrophic outliers ($CO\%$), the percentage of galaxies above $z =$ 1.5 which are outliers ($O\%>1.5$), the percentage of galaxies above $z =$ 1.5 which are catastrophic outliers ($CO\%>1.5$), the percentage of catastrophic outliers which are flagged ($CO_{\rm f}\%$), the percentage of catastrophic outliers above $z =$ 1.5 which are flagged ($CO_{\rm f}\%>1.5$), the percentage of galaxies above $z = 1.5$ which are non-flagged catastrophic outliers ($\%CO_{\rm nf}>1.5$), the percentage of non-outliers which are flagged ($NO_{\rm f}\%$), the percentage of non-outliers above $z =$ 1.5 which are flagged ($NO_{\rm f}\%>1.5$), the percentage of non-outliers above $z =$ 2.0 which are flagged ($NO_{\rm f}\%>2.0$), the percentage of galaxies above $z =$ 1.5 which are non-flagged non-outliers ($\%NO_{\rm nf}>1.5$), and the optimal parameters used for flagging ($p_{\rm f,min}$ and $\Delta z_{\rm peak,min}$) determined using the methods discussed in \S \ref{flagandopt}. Relevant to \S \ref{mod} we show results from determinations where some training sets have been modified to have a randomly undersampled redshift distribution. Relevant to \S \ref{pre} we show results from determinations where the dimmest 10\% of galaxies (in $i$-band flux) have been eliminated from the training set. Further details about these determinations are presented in the associated text section.}}
\end{table*}

We note that step (i) would require a dedicated campaign, rather than the use of previously assembled spectroscopic collections, to ensure the subset of galaxies for which spectroscopic redshifts are determined are as complete and representative as possible of the entire set of galaxies for which photometric redshifts are desired. The question could be raised as to whether it is realistic to expect that spectroscopic redshifts could be obtained for all 18,000 randomly selected galaxies. In order to explore this complication, we will assume (here for simplicity) that spectroscopic redshifts for the dimmest 10\% (in $i$-band) of the 18,000 selected galaxies cannot be determined. As examples, the resulting effect of this on the redshift distributions of our real galaxy test data sets is shown in Figure \ref{maglims}. In order to evaluate these simulated `spectro-magnitude-limited' cases we follow the example prescription, but start by randomly selecting 19,800 galaxies and then removing the dimmest 10\% of galaxies by $i$-band, then continuing the example prescription as described. Figures \ref{maglimflaggingperc} and \ref{maglimpercplots} compare the baseline and modified randomly undersampled cases for both COSMOS test data sets where the dimmest 10\% (in $i$-band) galaxies were removed from the training sets. We see the same improvements from the training set modification as for the non-spectro-magnitude-limited case, although the benefits are slightly less dramatic. Table \ref{tab2} shows several metrics for various cases of determinations with SPIDERz, with and without magnitude limits.

As in the non-spectro-magnitude-limited case, we find that using the COSMOSxSpecs test data set, even with a 10,000 galaxy evaluation set, we are not able to consistently find the same optimal flagging parameters as with the full evaluation set. However, in the case of COSMOS-reliable-$z$ we are able to consistently find optimal flagging parameters which are very close to those of the full evaluation set, but not as exact as with the non-spectro-magnitude-limited case. 

\clearpage

\section{Discussion} \label{disc}

This work presented two intertwined strategies for mitigating the effects of CO photo-$z$ estimates for large-scale surveys with a small number of photometric bands available: (i) using EPDFs to flag potential COs and (ii) modifying training sets to reduce high-redshift COs and flagged NOs. We evaluated these strategies using galaxy test data sets described in \S \ref{tdata} via analyses with SPIDERz, a custom support vector machine package for photo-$z$ estimation which naturally outputs an EPDF for each galaxy.  

In \S \ref{photozreal} we discussed the challenges that result from having only a limited number of optical bands with which to carry out photo-$z$ estimations, as well as those arising when a training set is not drawn from the same galaxy sample as the evaluation set for which photo-$z$s are desired.  Some quantifications of the degradation in performance when considering the reduction in bands and the use of so-called `mismatched'  training sets are shown in Table \ref{tab1}.

In \S \ref{flagandopt} we presented the strategy for flagging potential COs and showed some metrics of performance, indicating that a high percentage of COs can be flagged while simultaneously flagging a much lower percentage of NOs.  Several previous works \citep[e.g.][]{D08,MW08,ST13} have also explored the use of probability information to potentially reduce the numbers of outliers and COs in photo-$z$ estimates of large survey data sets, a process sometimes referred to in those works as ``cleaning.''  Of these works, \citet{MW08} and \citet{ST13} used the specific ODDS procedure as initially outlined in \citet{B00}, while \citet{D08} used a similar method. The present analysis of the potential of using EPDFs for flagging COs differs from those works in that: i) the previous works cut potential outliers based on the width of a $p(z)$ distribution, while this work explores a strategy based on identifying multiple, separated peaks of redshift probability and ii) this work deals with an empirical, machine learning approach to photometric redshift estimation rather than the template fitting approaches analyzed in those works. Those works use varying definitions of NOs, outliers, and COs, test data sets with varying redshift ranges, and other parameters making direct quantitative comparisons difficult. Nevertheless, we present some metrics for comparison to the results here, while emphasizing again that many parameters of these analyses differ from the present and from each other.

\cite{D08} report two results for identifying potential outliers: removing (synonymous with `flagging' in this work) 95\% of outliers and 34\% of galaxies total, and 90\% of outliers and 14\% of galaxies total. With about 3.5\% outliers for the first case and 1.8\% for the second, the authors of this paper remove over 30\% and over 12\% of NOs, respectively. While this is much higher than the current work, the definition of outliers used in \cite{D08} is less strict than in the current paper --- for a galaxy to be considered an outlier in that work, its estimation must be \textit{more incorrect} compared to the current work, meaning they define fewer galaxies as outliers than would be classified as such using the current work`s definition. This means that they are likely also removing galaxies which would fit the definition of outliers presented in this analysis, but which they are considering removed NOs. Applying the current analysis' definition of outliers to their results would likely lead to some reduction in the percentage of removed NO galaxies from the 30\% and 12\% figures, but it is likely that even with the modified definition of outliers the number would remain higher than in this analysis.

\cite{MW08} define a category of estimates redshifts whose error is $\Delta z < 0.5$ (hereafter `semi-non-outliers' for the sake of clarity). This definition sits between the definition of outlier and NO presented in the current analysis. With some calculations, one can determine that as little as around 9\% and as high as over 83\% of galaxies in that work (differing for various cases and metrics) would fall under the description of a `flagged semi-non-outlier' (comparable to flagged NOs presented in this work) --- i.e., galaxies which were identified as, or removed for, being outliers, and which were in reality semi-non-outliers. Since the definition of `semi-non-outliers' is less strict than the definition of NOs presented in this work, it is possible that a smaller fraction of those galaxies in \cite{MW08} whose estimations are more accurate are being lost, though it is not possible to be certain with the metrics presented. However, many of these metrics are well above those presented in the current analysis (sometimes 20 times higher), and so it seems safe to claim that for the majority of cases presented in \cite{MW08} more NOs are being lost than in the current work.

In \S \ref{mod} we showed that relatively simple modifications to the redshift distributions of training sets can improve both photo-$z$ accuracy and flagging accuracy at high redshifts.  It is interesting to consider that this work has explored two distinct ways in which a training set can be non-representative of an evaluation set.  In one case, as with the `mismatched' training sets discussed in \S \ref{photozreal} where the training set is drawn from a different underlying galaxy sample than the evaluation set, potentially containing systematically different distributions of galaxy types any given redshift interval, we see a major degradation in photo-$z$ determination accuracy.  In the other case, where the training set is drawn from the same underlying galaxy sample as the evaluation set, thus containing the same distributions of galaxy types in every redshift interval, but then modified by random undersampling as in \S \ref{mod} so as to have a different redshift distribution as the evaluation set but still the same distributions of galaxy type in every redshift interval, we see an improvement in photo-$z$ determination accuracy.  Thus there is quite some subtlety in empirical photo-$z$ determination regarding the applicability of the commonly-held notion that in machine learning the training set should be representative of the evaluation set.  

Based on these results, we presented an example prescription in \S \ref{pre} for applying these strategies to photo-$z$ estimation in a large-scale extragalactic survey to achieve more accurate photo-$z$ estimates and to mitigate the effects of CO estimates.  As discussed there, these methods should be operable in principle for any campaign in which photo-$z$s are determined with any empirical method which provides probability distribution over redshift information for each galaxy; however we emphasize the potential utility of SPIDERz in this regard.

\section*{Acknowledgements}

Based in part on observations made with the NASA/ESA Hubble Space Telescope, obtained from the Data Archive at the Space Telescope Science Institute, which is operated by the Association of Universities for Research in Astronomy, Inc., under NASA contract NAS 5-26555. Funding for the DEEP2 survey has been provided by NSF grants AST-0071048, AST-0071198, AST-0507428, and AST-0507483. Some of The data presented herein were obtained at the W. M. Keck Observatory, which is operated as a scientific partnership among the California Institute of Technology, the University of California and the National Aeronautics and Space Administration. The Observatory was made possible by the generous financial support of the W. M. Keck Foundation. The DEEP2 team and Keck Observatory acknowledge the very significant cultural role and reverence that the summit of Mauna Kea has always had within the indigenous Hawaiian community and appreciate the opportunity to conduct observations from this mountain.

\end{document}